\documentclass[sigconf]{acmart}

\AtBeginDocument{%
  \providecommand\BibTeX{{%
    \normalfont B\kern-0.5em{\scshape i\kern-0.25em b}\kern-0.8em\TeX}}}

\setcopyright{rightsretained}
\acmDOI{}

\acmConference[]{}{}{}
\acmBooktitle{}
\acmPrice{}
\acmISBN{}

\usepackage{algorithmic}
\usepackage{graphicx}
\usepackage{textcomp}
\usepackage{xcolor}
\usepackage[utf8]{inputenc}
\usepackage{graphicx}
\usepackage{textcomp}
\usepackage{xcolor}
\usepackage{listings}
\usepackage{enumitem}
\usepackage{algorithm}
\usepackage{algorithmic}
\usepackage{tikz}
\usepackage{url}
\usepackage{pgfplots}
\usepackage{tkz-kiviat} 
\usetikzlibrary{arrows}
\usepackage{subfig}
\usetikzlibrary{matrix}
\usetikzlibrary{positioning}
\usepackage{listings}
\usepackage{paralist}

\definecolor{mGreen}{rgb}{0,0.6,0}
\definecolor{mGray}{rgb}{0.5,0.5,0.5}
\definecolor{mPurple}{rgb}{0.58,0,0.82}
\definecolor{backgroundColour}{rgb}{0.95,0.95,0.92}

\lstdefinestyle{CStyle}{ 
    commentstyle=\color{mGreen},
    keywordstyle=\color{magenta},
    numberstyle=\tiny\color{mGray},
    stringstyle=\color{mPurple},
    basicstyle=\footnotesize,
    breakatwhitespace=false,         
    breaklines=true,                 
    captionpos=b,                    
    keepspaces=true,                 
    numbers=left,                    
    numbersep=5pt,                  
    showspaces=false,                
    showstringspaces=false,
    showtabs=false,                  
    tabsize=2,
    language=C,
}

\begin{document}

\title{Updatable Materialization of Approximate Constraints\\
}
\author{Steffen Kläbe}
\email{steffen.klaebe@tu-ilmenau.de}
\affiliation{%
  \institution{TU Ilmenau, Germany}
}

\author{Kai-Uwe Sattler}
\email{kus@tu-ilmenau.de}
\affiliation{%
\institution{TU Ilmenau, Germany}
}

\author{Stephan Baumann}
\email{stephan.baumann@actian.com}
\affiliation{%
\institution{Actian Germany GmbH}
}

\begin{abstract}
Modern big data applications integrate data from various sources. As a result, these datasets may not satisfy perfect constraints, leading to sparse schema information and non-optimal query performance. The existing approach of PatchIndexes enable the definition of approximate constraints and improve query performance by exploiting the materialized constraint information. As real world data warehouse workloads are often not limited to read-only queries, we enhance the PatchIndex structure towards an update-conscious design in this paper. Therefore, we present a sharded bitmap as the underlying data structure which offers efficient update operations, and describe approaches to maintain approximate constraints under updates, avoiding index recomputations and full table scans. In our evaluation, we prove that PatchIndexes significantly impact query performance while achieving lightweight update support.

\end{abstract}

\maketitle

\section{Introduction}
In classical database theory, database schemas are well-defined and contain a set of database constraints, like primary and foreign keys, uniqueness definitions or sort keys, which offer two major advantages. First, defining constraints ensures data integrity, which is handled automatically by the database management system (DBMS) whenever updates on the data occur. Second, constraint definitions are used in query optimization to accelerate queries containing joins, aggregations or sort operations. Here, query optimizers can benefit from the constraint information by choosing optimal physical operators or optimal query plans.

Real world cloud applications often do not follow this classical theory of a well-defined database schema and contain sparse schema information. This is reinforced by the lack of a database administrator who tunes the database continuously. In general, big data applications often contain unclean data for different reasons like integrating different sources with heterogeneous schemas or real world anomalies like duplicate names, shared addresses or telephone numbers. Thus, perfect constraints may not exist in these datasets. Nevertheless, they may contain approximate constraints, which are constraints that hold for all tuples except a small set of exceptions (captured as ``exception rate''). These exceptions prohibit the definition of a constraint, resulting in the loss of valuable information. 
Examples for approximate constraints are ``nearly unique columns'' (NUC) and ``nearly sorted columns'' (NSC), whose definition we introduced in \cite{klabe_patchindex_2020}. 

\begin{figure}
\begin{tikzpicture}
\begin{axis}[
    ymin=0,ymax=16,
    ybar,
    width=8cm,
    height=4.5cm,
    tick pos=left,
    bar width=0.2cm,
    xmin=0,xmax=100,
    xmajorgrids,
    xtick align=inside,
    ytick distance={2},
    legend style={at={(0.5,-0.15)},
      anchor=north,legend columns=-1},
    ylabel={\#columns},
    xlabel={Amount of tuples that match the constraint [\%]},
    xtick={0,20,40,60,80,100},
    legend style={at={(1,1)},anchor=north east},
    legend columns=1,
    legend cell align={left}
    ]
\addplot coordinates {(10,504) (30,5) (50,1)(70,5)(90,4)};
\addlegendentry{USCensus\_1 (NSC)};
\addplot coordinates {(10,14) (30,0) (50,2)(70,0)(90,4)};
\addlegendentry{IGlocations2\_1 (NUC)};
\addplot coordinates {(10,14) (30,3) (50,2)(70,1)(90,7)};
\addlegendentry{IUBlibrary\_1 (NUC)};
\end{axis}
\end{tikzpicture}
\vspace{-1em}
\caption{Histogram over approximate constraint columns in PublicBI datasets}
\label{publicbi_histo}
\vspace{-2em}
\end{figure}
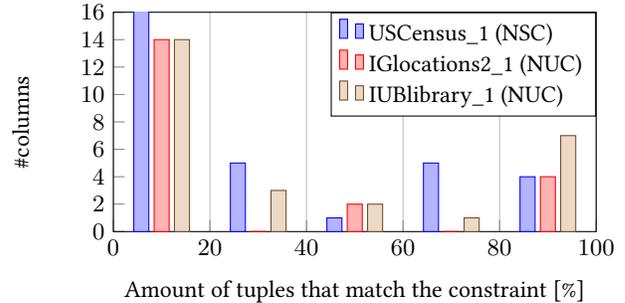

Investigations on real world datasets prove the existence of approximate constraints and encourage the need to handle them by the DBMS. The PublicBI benchmark \cite{vogelsgesang_get_2018} is a collection of Tableau workbooks and allows experimental evaluation against real user datasets with their common properties, e.g. many string columns, many NULL values or the absence of constraint definitions. Some of these datasets showed a large number of approximate constraints, like the \emph{USCensus\_1}, the \emph{IGlocations2\_1} and the \emph{IUBlibrary\_1} workbook. The histogram shown in Figure~\ref{publicbi_histo} represents the distribution of approximate constraint columns for these datasets. The \emph{USCensus\_1} workbook contains over 500 columns, from which 15 columns match an approximate sorting constraint. Nine columns match the sorting constraint with over 60\% of their tuples. The \emph{IGlocations2\_1} and the \emph{IUBlibrary\_1} workbooks contain a small number of columns, from which a relatively large amount follow an approximate uniqueness constraint. Many of these columns are nearly perfectly unique.

In \cite{klabe_patchindex_2020} we proposed the concept of PatchIndexes, which allows the definition of approximate constraints and enables database systems to benefit from the constraint information in query execution by materializing exceptions and handling them separately. We provided constraint discovery mechanisms for NUC and NSC and compared different design approaches. The evaluation showed the positive impact of the concept on query performance even for high exception rates.
Real world OLAP applications often differ from the theoretical assumption of a read-only data warehouse with periodical bulk updates, as data freshness becomes more important in decision-making processes and HTAP workloads. Consequently, update support is of significant performance in these applications. As the main issue with the materialization of information is the update support, we further enhance the PatchIndex feature in this paper in the direction of a generic and updatable materialization approach for approximate constraints. We explicitly exclude the discovery of these approximate constraints, as already discussed in \cite{klabe_patchindex_2020}, and focus on an update-conscious index design and query support. Our main contributions of this paper are: 

\begin{compactitem}
\item We design a sharded bitmap data structure as the underlying concept of the PatchIndex, offering efficient insert, modify and delete operations.
\item As delete operations are the major challenge for the underlying sharded bitmap data structure, we present a parallel and vectorized approach for delete support.
\item We provide lightweight mechanisms to discover deltas of the materialized PatchIndex information in case of update queries. Consequently, constraints are allowed to become approximate over time, even if they were perfect at the time of definition. This reduces the abort frequency of update operations compared to perfect constraints.
\end{compactitem}

The remainder of the paper is organized as follows: We discuss related work in Section~\ref{related} and recap the background of PatchIndexes in Section~\ref{background}. In Section~\ref{design}, we present the sharded bitmap structure as an update-conscious index design, before describing update handling mechanisms in Section~\ref{update_handling}. We evaluate our approach in Section~\ref{eval} before concluding in Section~\ref{conclusion}.
\section{Related work}\label{related}
The materialization of information is a widely investigated topic in the database research field, particularly in the context of materialized views. While significantly impacting query performance \cite{blakeley_join_1990}, the major drawback of materialized views is their update support. For the subset of select-project-join (SPJ) views there are efficient ways to support table updates and keeping views consistent \cite{blakeley_updating_1989, blakeley_efficiently_1986, gupta_using_1995, tompa_maintaining_1988}. They typically try to avoid base table access by determining update sets and performing differential update operations. Besides that, materialized information should be chosen based on the expected benefit, so intermediate results of frequent queries are a good candidate for materialization. A benefit-based approach for choosing intermediate results is given in \cite{nagel_recycling_2013}, while \cite{hagedorn_cost-based_2018} also includes the probability of a materialized result to be usable in future queries to the cost model. Nevertheless, these approaches do not take updatability into account. 

Besides materialized views, there are several specialized materialization approaches like SortKeys, materializing a certain tuple order, or JoinIndexes \cite{valduriez_join_1987} and Bitwise Dimensional Co-Clustering (BDCC) \cite{baumann_bitwise_2016} for join materialization. While JoinIndexes materialize foreign key joins by maintaining an index to the join partner as an additional table column, BDCC physically co-locates join partners of different tables in distributed databases.
Our approach materializes aggregate, join and sort results directly at the table level, which makes it usable for a wide range of queries while still providing efficient update support. Particularly updating uniqueness constraints, which are the results of aggregations, differs from the general approaches for SPJ views.

Research on approximate constraints evolved from the field of constraint discovery. For the uniqueness constraint, approaches for unique column combinations (UCCs) \cite{abedjan_profiling_2015} and respective discovery algorithms \cite{heise_scalable_2013, papenbrock_hybrid_2017} are introduced to handle perfect constraints. In order to also handle approximate uniqueness constraints, the concepts of ``possible'' and ``certain'' keys are presented in \cite{kohler_possible_2015} to enforce constraints by replacing violating tuples. Besides that, embedded uniqueness constraints (eUC) \cite{wei_discovery_2019} separate uniqueness from completeness to enforce the uniqueness constraint only on a subset of tuples. Recent publications \cite{pena_discovery_2019, livshits_approximate_2020} also cover discovery approaches for approximate denial constraints, a more general class of data-specific constraints. As these approaches mainly focus on the discovery of approximate constraints, we integrate approximate constraints into query execution in an updatable way with our paper. We therefore combine the concept of approximate constraints with the concept of materialization.

We base our approach on the concept of patch processing, which is commonly used in compression. The PFOR, PFOR-DELTA and PDICT compression schemes \cite{zukowski_super-scalar_2006} show robustness against outliers by handling exceptions to certain distributions separately. Furthermore, white-box compression \cite{ghita_white-box_2020} learns distributions and properties of data in order to choose appropriate compressing schemes. Here, tuples that do not follow a certain behaviour can be compressed using another compression scheme than remaining tuples, significantly improving compression rates as a result. These approaches modify the way data is physically stored to handle exceptions, which significantly differs from our approach.  By not changing the physical order of data, our concept of PatchIndexes allows multiple approximate constraint definitions per table, so for example the definition of multiple approximate SortKeys.

\section{Background}\label{background}

In \cite{klabe_patchindex_2020} we introduced the concept of PatchIndexes, enabling database systems to define and materialize approximate constraints. In this section, we briefly recap the basic problem definition, the approaches for the PatchIndex design and its integration into query optimization.  
\vspace{-0.25em}
\subsection{Problem Definition}
\vspace{-0.25em}
An approximate constraint is a constraint that is satisfied by all tuples except a set of exceptions. We identify these exceptions using their rowIDs and refer to a tuple violating a certain constraint as a \emph{patch}. Hence, the \emph{set of patches} $P_c$ is the set of all rowIDs violating a certain constraint in a column $c$. The ratio  between the number of exceptions and the total number of tuples is defined as the \emph{exception rate} $e$. As examples, we introduced ``nearly unique columns'' (NUC) and ``nearly sorted columns'' (NSC) and provided discovery mechanisms determining a minimal set of patches to the respective constraints.

\vspace{-0.25em}
\subsection{Index Design}\label{back_design}
\vspace{-0.25em}
The main task of the PatchIndex structure is the efficient representation of the set of patches for a given constraint. We investigated two basic approaches for the design. While the identifier-based approach maintains a list of the 64\;bit tuple identifiers of patches and is therefore the sparse way of storing, the bitmap-based approach is the dense way of storing and contains a single bit for each tuple of the indexed column, indicating whether it is an exception or not. The choice between both approaches is a trade-off in terms of memory consumption, with the bitmap-based approach consuming less memory for cases with $e \geq \frac{1}{64} = 1.56\%$. Data partitioning is transparent for PatchIndexes, as a separate index is created for each partition. Constraint discovery, index creation and query processing are performed partition-locally and in parallel.

\vspace{-0.25em}
\subsection{Query Processing}\label{back_queryproc}
\vspace{-0.25em}
The main idea to integrate PatchIndexes into query execution is the PatchIndex scan, merging the PatchIndex information on-the-fly with the dataflow and splitting it into a flow of tuples satisfying the constraint and a flow of exceptions, and work on both dataflows separately. The PatchIndex scan is realized using an additional selection operator on top of a scan operator, merging the PatchIndex information on-the-fly to the dataflow according to selection modes \emph{exclude\_patches} and \emph{use\_patches}. The key for query optimization using PatchIndexes is cloning query subtrees for data and patches and optimizing both subtrees separately. Typically, we can exploit the PatchIndex information in the dataflow that excluded the patches and achieve a speedup in this subtree, as a constraint is fulfilled here. Finally, both subtrees are combined again to ensure transparency for the remaining query execution tree.

\begin{figure}[htbp]
\centering
\vspace{-1em}
\includegraphics[width=0.45\textwidth]{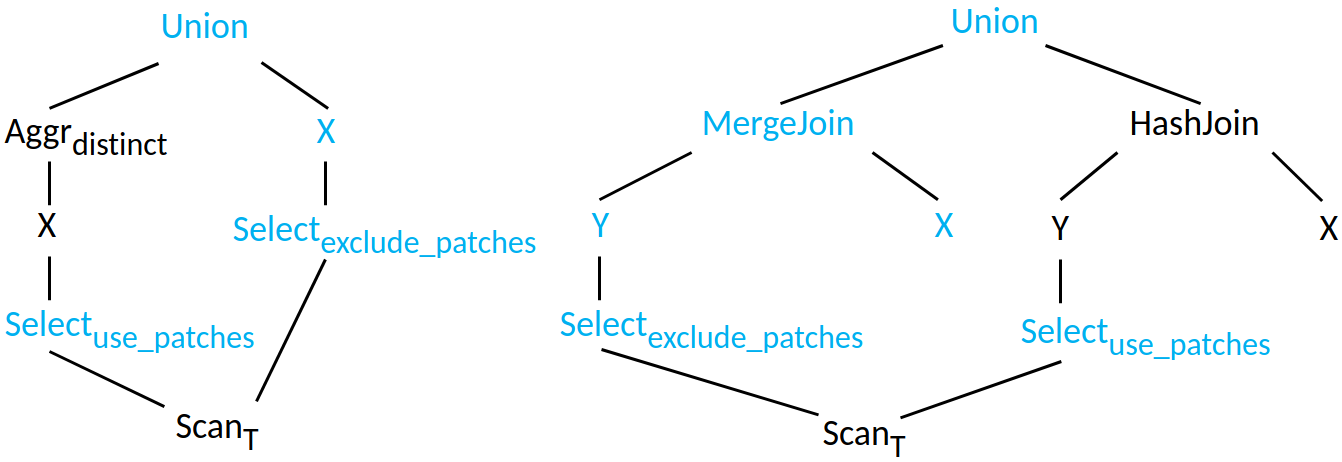}
\vspace{-0.5em}
\caption{Query plans for distinct (left) and join (right) queries before (black operators) and after PatchIndex optimization with the newly inserted operators highlighted in blue.}
\vspace{-0.5em}
\label{query_plans}
\end{figure}

PatchIndexes can be integrated into the optimization of distinct, join and sort queries. For distinct queries, we can exploit the information that tuples are unique in a NUC after excluding the patches, so the most expensive aggregation operator to compute the distinct values can be dropped from this subtree as shown on the left side of Figure~\ref{query_plans}. Here ``X'' is an arbitrary subtree that does not contain any joins or aggregations. In case of a query that contains a grouping, this approach can also be applied by decoupling the aggregation for grouping from the distinct aggregation.

Knowing about the sorting of NSC in join queries, we can replace the generic HashJoin operator with the faster MergeJoin operator in the subtree that excluded patches, like shown in the right part of Figure~\ref{query_plans}. This optimization requires the subtree ``X'' to be sorted on the join key, which is a frequent case in joins between fact tables and dimension tables in data warehouse applications. While subtree ``Y'' must preserve the tuple order and is therefore not allowed to contain aggregations, both ``X'' and ``Y'' may contain join operators that are order preserving, so e.g. being the probe side of a HashJoin. Under these requirements the PatchIndex information can be propagated through a query tree.
In order to further optimize this approach, the result of subtree ``X'' is buffered instead of computed twice. Additionally, join sides can be swapped to further accelerate the HashJoin. As the number of patches is known during optimization time, we can choose the build side of the HashJoin as the one with the lowest cardinality to improve the time and space for hash table building. Building the hash table on the patches is often the best decision as the number of patches is typically small. The join sides of the MergeJoin are chosen similarly to ensure the combinability of both subtrees. As the MergeJoin operator is not sensitive to join sides, this has no impact on performance.

Besides the join operator, the information about a NSC can also be used to accelerate sort queries, as we already know that a major part of the data is already sorted. Here we follow an approach similar to distinct queries, so the query plan is equal to the left plan of Figure~\ref{query_plans} exchanging the aggregation with the sort operator. If there is a PatchIndex defined on the sort column using the same ordering as the sort operator, the sort operator becomes obsolete in the subtree that excludes patches, as these tuples are already known to be sorted. The sort operation only needs to be performed on the patches, which is intended to be the minor part of the data. In order to preserve the sort order, the results of both subtrees are combined using a Merge operator instead of a Union operator. 
Our evaluation in \cite{klabe_patchindex_2020} showed that PatchIndexes significantly increase query performance for different queries and even high exception rates, resulting in the choice for the bitmap-based design approach due to it's constant memory consumption.

\vspace{-0.25em}
\subsection{Recovery}
\vspace{-0.25em}
PatchIndexes are currently designed as main memory data structures. In order to keep the database log as slim as possible, we avoid logging the actual  PatchIndex information, so PatchIndexes are recreated after a system shutdown or failure. Alternatively, the PatchIndex information can be persisted to disk as a checkpoint in combination with the logging of succeeding update operations after persisting the index state.

\vspace{-0.25em}
\subsection{Cost model}\label{cost_model}
\vspace{-0.25em}
The use cases for query optimization presented in the preceeding subsections can be easily integrated into the cost models of arbitrary query optimizers, as cardinalities and operator output estimates are known during optimization time and we use ordinary query operators for the optimization, except the newly introduced selection modes for the PatchIndex scan. The costs for query trees (build and execution costs) produced by the PatchIndex optimizations can therefore be estimated by query optimizers and if the estimated costs are smaller than the costs of the subtree before the optimization, the PatchIndex should be used for the query. The seletion operators with modes \emph{exclude\_patches} and \emph{use\_patches} merge the PatchIndex information on-the-fly to the output dataflow of the scan operator. This is particularly independent from the data types of the input data, as the decision of passing or dropping a tuple is based on a tuple's rowID. As a consequence, the operator's overhead is fixed for every tuple. In our experiments, the selection operators for both selection modes took a minor part of query runtimes (typically below 1\%).

\section{Update-conscious Index Design}\label{design}

In database systems, there are three types of table updates, namely insert, modify and delete operations. With the bitmap-based approach described in Section~\ref{back_design}, inserts and modifies can be handled efficiently. For inserts, reallocating/resizing the bitmap and setting single bits is sufficient, while for modifies single bits need to be changed. 
Therefore we mainly focus on delete operations, which are the main challenge for the bitmap structure, as it needs to be ensured that read access to specific bits remains efficient after update operations. Although it can be realized by shifting the bitmap towards the deleted position, this potentially shifts large amounts of memory and therefore results in poor performance. In this section, we present our approach of a sharded bitmap that efficiently supports delete operations while keeping additional memory consumption low. We base our design approach on existing update-conscious bitmap approaches and extend them with parallel and vectorized delete and bulk delete support.
\vspace{-0.25em}
\subsection{Sharded bitmap design}
\vspace{-0.25em}

Update Conscious Bitmap Indices \cite{canahuate_update_2007} faced the problem of degrading read performance under updates \cite{athanassoulis_upbit_2016}. HICAMP \cite{wang_hicamp_2014} and UpBit \cite{athanassoulis_upbit_2016} proposed different solutions to this problem for multi-dimensional bitmap indexes. While HICAMP divides bitmaps into slices and maintains access to them over a directed, acyclic graph, UpBit uses delta-structures for update maintenance and ``fence pointers'' for fast access to arbitrary bit positions. Both approaches share the concept of slicing, either explicitly or implicitly over pointers, to keep updates local and allow parallel operations on the bitmap.

\begin{figure}[htbp]
\centering
\includegraphics[width=0.475\textwidth]{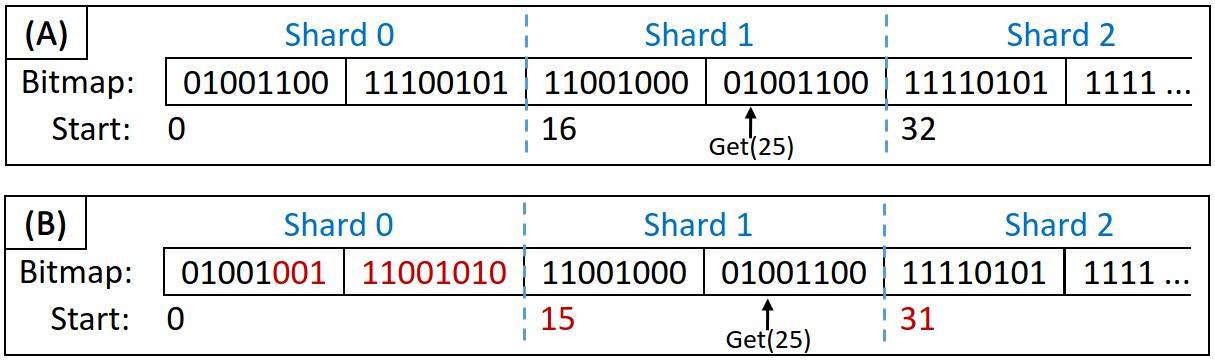}
\vspace{-2em}
\caption{Sharded bitmap design before (A) and after (B) deleting the bit at position 5.}
\vspace{-1em}
\label{sharded_bitmap}
\end{figure}

Similar to UpBit, in our sharded bitmap design we implicitly divide bitmaps into virtual shards in order to keep delete operations local and to enable parallelism in update operations. The upper part of Figure~\ref{sharded_bitmap} shows the design concept of the data structure. The bitmap is realized using an array of addressable elements. While these elements are 64\;bit types
in our implementation, we reduced the size to 8\; bits in Figure~\ref{sharded_bitmap} for clarification. A virtual shard then consists of multiple addressable elements and a single additional integer value indicating the index of the first bit in a shard, similar to UpBit's fence pointers. Besides keeping delete operations local, this approach also facilitates fine-grained locking and logging for efficient concurrency control as described in Section~\ref{concurrency}.

\vspace{-0.25em}
\subsection{Sharded bitmap operations}
\vspace{-0.25em}
\subsubsection{Bit access}\label{bit_access}
In an ordinary bitmap, single bit access methods \emph{set()}, \emph{get()} and \emph{unset()} are performed in two steps. First, the position of the bit is calculated by getting the addressable element using a division of the position by the size of an addressable element (realized using a bit shift) and getting the position within an element using a bitwise AND operation with a bitmask. In a second step, the bit is changed or returned using another bitwise operation.

In order to access a single bit in the sharded bitmap structure, we first compute the shard containing the bit. This can be efficiently realized using a bit shift and additional comparisons with the start values of the upcoming shards. The additional checks are necessary as a bit may be contained in a subsequent shard due to previous delete operations. Then the bit can be accessed using the position relative to the start value of the shard in a similar way than the ordinary bitmap. This way, accessing a single bit in the sharded bitmap is only slightly slower than in ordinary bitmaps. 

\subsubsection{Delete}
Deleting a single bit in the sharded bitmap is divided into three steps: 
\begin{compactenum}[(a)]
\item Determine the position of the bit like described in Section~\ref{bit_access}.
\item Shift all subsequent bits within the shard by one position towards the deleted bit.
\item Decrement the start values of all subsequent shards.
\end{compactenum}

As an example, the result of deleting the bit at position~5 is shown in the lower part of Figure~\ref{sharded_bitmap} with the changed elements highlighted in red color. Note that after deleting a bit, the values of subsequent elements change, which is the desired semantic of the delete operation. In the example, the bit at position ~25 after the delete operation is the bit at position~26 before the delete. Compared to ordinary bitmaps, steps~(b) and (c) are tradeoffs, as we limit the impact of a delete operation to a single shard in step~(b), but need additional effort afterwards in step~(c) to adjust metadata. Therefore, the choice of the shard size is crucial for the data structure.

Step~(b) is the main challenge of this approach as it involves cross-element bit shifts, which can be realized by a sequence of bit masking and shifting operations. In order to further accelerate this step, we designed a vectorized cross-element shifting algorithm that uses Advanced Vector Extensions Version 2 (AVX2) intrinsics and is based on shifting, masking and permutation intrinsics to enable data exchange between AVX lanes. The algorithm that is used inside the loop over the data of a shard is shown in Listing~\ref{lst_avx} (assuming pre-defined constant bit masks for shifting and masking).

\makeatletter
\lst@CCPutMacro
    \lst@ProcessOther {"2A}{%
         {\raisebox{1pt}{*}}%
         }
    \@empty\z@\@empty
\makeatother  
\lstset{numbers=left,xleftmargin=1.2em, belowskip=-2.5em}
\begin{lstlisting}[style=CStyle, caption={Cross-element bit shift using AVX2},label=lst_avx, float]
// Load data to axv vector
__m256i x = _mm256_loadu_si256((__m256i*)data);
// Get last bit of each element 
__m256i y = _mm256_and_si256 (x, bit_mask);
// First bit will be last bit of prev. element 
y = _mm256_sllv_epi64(x, shift_mask63);
// Save element at pos 0 for next iteration
bits = _mm256_blend_epi32(bits, y, 0x03);
// Rotation Mask: (3, 3, 2, 1) is 1111001
__m256i rotated=_mm256_permute4x64_epi64(y, 0xF9);
// Copy element from last iteration to pos 3
rotated = _mm256_blend_epi32(rotated, bits, 0xC0);
// Saved element to pos 3 for next iteration 
bits = _mm256_permute4x64_epi64(bits, 0x24);
// Shift data 
x = _mm256_srlv_epi64(x, shift_mask1);
// Insert bit from next element
x = _mm256_or_si256 (x, rotated);
// Store back to data
_mm256_storeu_si256((__m256i*)data, x);
\end{lstlisting}

\subsubsection{Bulk delete}\label{bulk_delete}
\begin{figure}[htbp]
\centering
\vspace{-1em}
\includegraphics[width=0.48\textwidth]{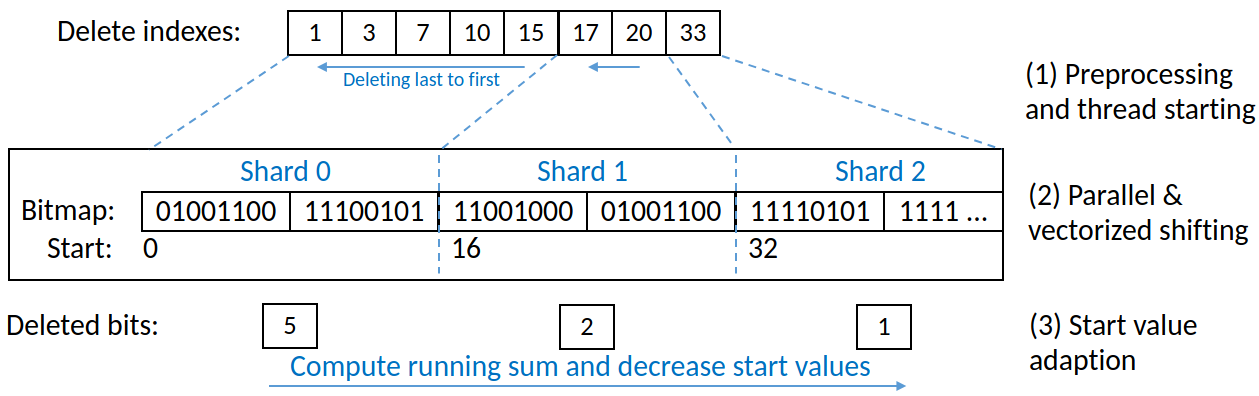}
\vspace{-2em}
\caption{Parallel \& vectorized bulk delete operation}
\label{img_bulk_delete}
\end{figure}

In order to reduce the effort of changing the start value array and to exploit the opportinities of the sharding approach for parallelism, we introduce a bulk delete operation to the sharded bitmap structure. The basic concept of the bulk delete operation is shown in Figure~\ref{img_bulk_delete}. After a preprocessing step to determine the shards that belong to the elements, step~(b) of shifting within the single shards can be performed in parallel using threads, as bitshifts remain local to the shards due to the design of the data structure. A thread is hereby created for each shard that contains indexes to be deleted, so the total number of threads depends on the number of shards and the location of deleted positions. The actual bitshift is again realized using the vectorized algorithm shown in Listing~\ref{lst_avx}. At the end of the operation, all start values are adapted in a single array traversal by holding a running sum over deleted bits of all preceeding groups. This way, the vectorized and parallelized bulk delete operation minimizes the additional effort of step~(c) while further accelerating step~(b).

The bulk delete operation is order sensitive, as deleting a bit within a shard shifts subsequent bits, changing their position. Consequently, positions of bits that are intended to be deleted are not correct if bits at smaller positions were deleted before. To overcome this, delete operations are performed in descending order, starting from the largest position. In our implementation, the PatchIndex structure buffers the rowIDs of tuples that should be deleted and performs a bulk delete operation at the end of the query. It thereby ensures the order of the rowIDs to enable an efficient bulk delete.

\subsubsection{Condense}
With each single delete operation, a bit at the end of a shard gets lost, as the subsequent bit from the following shard is not shifted to this position. This is the main drawback of the sharded bitmap design and follows from limiting the impact of a delete operation to a single shard. In order to overcome this drawback, we introduce a condense operation to the sharded bitmap design, which shifts the elements of the bitmap between shards and resets the utilization of the structure as a result. The condense operation is realized using a single traversal over the bitmap. For each shard, data of subsequent shards is shifted to the bits that were lost due to delete operations and the start values are adapted accordingly. Condensing can be triggered manually or automatically by monitoring the utilization of the bitmap and triggering once a certain threshold is reached.

\section{Handling Update Queries}\label{update_handling}

Handling table update operations like inserts, deletes or modifies is the most significant problem of materialization. As table updates were not present at the time a materialization was computed, it reaches an inconsistent state whenever an update occurs. The common way to handle this inconsistency is refreshing, often leading to an expensive recomputation. For some use cases, minor inconsistencies might be acceptable and refresh cycles can be chosen quite loose, while other applications rely on consistent and up-to-date data and therefore demand for very tight or just-in-time refresh cycles.

\begin{table*}
\centering

\begin{tabular}{p{1.2cm}p{6.8cm}p{6.8cm}}
 \hline
  & \textbf{Nearly unique column} & \textbf{Nearly sorted column} \\
 \hline

 \textbf{Insert} & Scan inserted tuples, join them with the table, merge the results with the existing patches. & Determine a new sorted subsequence extending the already existing one.\\
 \hline
 \textbf{Delete} & Drop tracking information about deleted tuples. & Drop tracking information about deleted tuples.\\
 \hline
 \textbf{Modify} & Scan modified tuples, join them with the table, merge the results with the existing patches. & Merge all modified tuples with the existing patches.\\
  \hline
\end{tabular}

\caption{Design principles for update operations}
\label{update_handling_ideas}
\vspace{-2em}
\end{table*}

On the contrary, PatchIndexes are designed to efficiently support table update operations. We avoid getting inconsistent states by handling updates immediately after they occur. In order to perform these updates efficiently, the design of the update handling mechanisms is driven by the goal of avoiding an index recomputation and avoiding a full table scan while preserving the invariant of holding all exceptions to a given constraint. 
The basic ideas for handling inserts, deletes and modifies for the specific constraints are summarized in Table~\ref{update_handling_ideas}.

For the design of the update handling mechanisms we utilized the following technologies that are commonly used in analytical, column-store based DBMS:

\begin{compactitem}
\item \textbf{Delta structures:} Delta structures are used to maintain table updates in-memory instead of writing them to disk, which is a typical approach for read-optimized column-stores. We used the concept of Postitional Delta Trees \cite{heman_positional_2010}, which was already integrated in our test system. PDTs are not coupled with but queried by PatchIndexes.
\item \textbf{Summary tables:} Summary tables are widely used in existing DBMS to collect certain information about columns. As an example, we used small materialized aggregates \cite{moerkotte_small_1998}, also called Minmax indexes, which materialize minimum and maximum values for buckets of tuples, enabling data pruning during scans by the evaluation of e.g. selection predicates.
\item \textbf{Range propagation:} Range propagation (or selection propagation in \cite{baumann_bitwise_2016}) extends the use of minmax indexes to push scan ranges across join operators. While static range propagation enables pushing scan ranges across joins during query build phase, dynamic range propagation dynamically generates scan ranges during query execution, e.g. during the build phase of HashJoins.
\item \textbf{Intermediate result caching:} Caching can be used within queries to avoid expensive recomputations. In our further considerations, we use the Reuse operator to encapsulate this behaviour. Intermediate results are materialized in main memory by the ReuseCache operator and read again from the ReuseLoad operator.
\end{compactitem}

\vspace{-0.25em}
\subsection{Insert}\label{insert_handling}
\begin{figure}[htbp]
\centering
\vspace{-1em}
\includegraphics[width=0.3\textwidth]{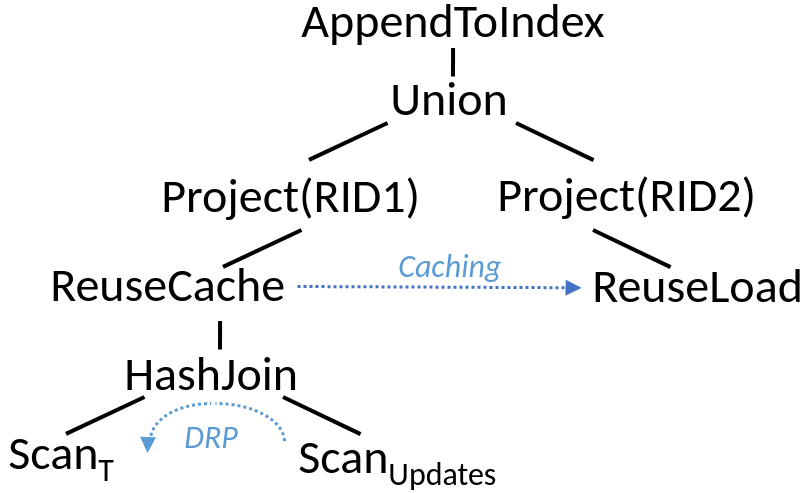}
\vspace{-0.5em}
\caption{Insert handling query on table~T with dynamic range propagation (DRP) and intermediate result caching}
\vspace{-1em}
\label{qet_insert_handling}
\end{figure}

\vspace{-0.25em}
Handling insert operations in the PatchIndex translates to answering the question which tuples have to be added to the existing patches. For the uniqueness constraint, this is a difficult question, as this constraint relies on a global view of the table. Inserting a single tuple might produce a collision with another tuple that was already in the table and had a unique value in the indexed column before. As we need to keep track of all occurences of non-unique values to ensure correctness, we perform a join query after the insert operation, joining the inserted tuples with the actual table (including inserted values, as duplicates might also occur in the inserts) like shown in Figure~\ref{qet_insert_handling}. Afterwards we project rowIDs of both join sides using intermediate result caching. The rowIDs are then merged into the existing patches. This way, we preserve the capability of holding the minimum set of patches to make the table unique when excluding them, while also avoiding a recomputation of the full index. In order to avoid the full table scan we utilize dynamic range propagation. After the build phase of the join operator is finished, scanning the full table is reduced to only the blocks that contain potential join partners. This significantly reduces I/O overhead and is therefore a major improvement of the insert handling mechanism.

For the sorting constraint, we focus on a local view on the inserted tuples instead of a global view on the whole table by extending the already existing sorted subsequence with inserted values instead of recomputing a globally longest sorted subsequence. The PatchIndex keeps track of the last value of the subsequence (i.e. the largest value of an ascending sort order or the smallest value of a descending sort order). During execution of the insert query, the PatchIndex temporaryly stores inserted values and starts computing a longest sorted subsequence with all values being larger/smaller than the last value of the existing sorted sequence. We hereby utilize the same longest sorted subsequence algorithm \cite{fredman_computing_1975} as used in the discovery of a NSC in \cite{klabe_patchindex_2020}. The rowIDs of tuples that are not included in this extending sorted subsequence are added to the set of patches. With this mechanism, we probably lose the optimality of the index, namely to keep track of the longest sorted subsequence. The reason for that is the fact that combining two sorted sequences that have a maximum length for two parts of the data might not form a longest sorted sequence for the combination of the two parts. An example for this can be easily constructed, e.g. the table holding values (1,~2,~10) and inserted values (3,~4). Nevertheless this does not lead to wrong query results when applying the PatchIndex information like described in Section~\ref{back_queryproc} and should be negligible for the typical use cases. However, monitoring the exception rate and triggering a global recomputation once a certain threshold is reached is a possible solution for that.

From an implementation point of view, scanning the inserted values is realized by scanning the PDTs of the current query. Furthermore, merging the determined results with the existing patches translates to merging the two lists for the identifier-based approach or reallocating the bitmap and setting the respective bits for the new patches in the bitmap-based approach.

\vspace{-0.25em}
\subsection{Modify}
\vspace{-0.25em}
For the uniqueness constraint, handling modify operations is done similarly to insert handling described in Section~\ref{insert_handling}. For the sorting constraint, all modified tuples have to be added to the set of patches, as modifying values might destroy the sorting of the computed subsequence. The only difference in the actual realization is that reallocating the bitmap for the bitmap-based approach becomes obsolete, as the table cardinality does not change during modify operations.

\vspace{-0.25em}
\subsection{Delete}
\vspace{-0.25em}
Deletions are handled by the PatchIndex by dropping the information about deleted tuples from the patch information without taking a global picture of the table into account. For both uniqueness and sorting constraint, dropping values from the table does not violate the constraint. For the uniqueness constraint, we might lose the optimality this way, as a value that was not unique before the delete operation might become unique afterwards if the other occurrences of the value was deleted. Nevertheless, it would remain in the patches to keep the delete handling mechanism simple and would obviously not lead to wrong query results when applying the PatchIndex to queries. The same holds for the sorting constraint, probably losing optimality when deleting values from the existing longest subsequence. As described in Section~\ref{insert_handling}, monitoring and recomputing the index is a possible solution if this becomes a problem. From an implementation point of view, deleting values raises the problem that rowIDs of subsequent tuples decrease for each deleted tuple. This is considered by the bulk delete operation of the sharded bitmap data structure as described in Section~\ref{bulk_delete}. For the identifier-based approach we keep track of the number of deleted tuples with smaller rowIDs and decrement each identifier while going through the list of patches.

\vspace{-0.25em}
\subsection{Concurrency} \label{concurrency} 
\vspace{-0.25em}
In terms of concurrency control, PatchIndexes seamlessly integrate into a system's snapshot isolation mechanism \cite{cahill_serializable_2008}. Although snapshot isolation is a very coarse-grained way to achieve serializability of transactions, it is often sufficient for read-optimized DBMS. Nevertheless, PatchIndexes offer opportunities for a more fine-grained concurrency control due to the underlying sharded bitmap data structure. As shards are independent from each other, fine-grained locking can be used to avoid concurrent access without locking the whole data structure. Adapting the start values of a sharded bitmap produces no conflicts, as the only operation that adapts the start values is the delete operation which uses decrement operations. Concurrent decrements are no conflicts, as different orders of executing a series of decrements produce the same result.

\vspace{-0.25em}
\subsection{Expandability}
\vspace{-0.25em}
In the preceeding discussions we presented the PatchIndex approach for NUC and NSC. However, PatchIndexes are not limited to these constraints due to the generic design of the data structure. Different constraints could be easily integrated by (1) implementing the PatchIndex interface for constraint-specific initial filling, insert, modify and delete support and (2) adding an optimizer rule to automatically rewrite query plans using the existing PatchIndex scan in order to exploit the approximate constraint information. We leave the investigation on different approximate constraints open for future work.
\section{Evaluation}\label{eval}
In this section, we evaluate our solution in different experiments to prove the performance impact of PatchIndexes and its updatability. We therefore show experiments on different integration levels, starting from the lowest level of the underlying data structure. Here, we present a comparison between the sharded bitmap approach and ordinary bitmaps and determine the optimal shard size. Second, we present a set of PatchIndex microbenchmarks for a fine-grained evaluation of aspects like performance impact, creation time or update operations. Afterwards we show the impact of PatchIndexes on query performance using a query subset of the well-known TPC-H benchmark. For the evaluation, we integrated PatchIndexes into the Actian Vector 6.0 commercial DBMS, which is built on the X100/Vectorwise \cite{boncz_monetdbx100_2005} analytical database engine. The system runs on a machine consisting of two Intel(R) Xeon(R) CPU E5-2680 v3 with 2.50\;GHz, offering 12 physical cores each, 256\;GB DDR4
RAM and 12\;TB SSD. For all measured results, we used queries on hot data, which means that data resides in the in-memory buffers of the system. This way, we reduce the I/O impact and focus on the pure query execution time. Additionally, we did not trigger any sharded bitmap condense operations but started every single experiment with a freshly build index structure for comparability reasons.

In the evaluation, we compare the generic PatchIndex approach against different specialized materialization approaches, namely materialized views, SortKeys and JoinIndexes. For distinct queries, we used materialized views as a comparison, which is a widely used technique in database systems to pre-compute partial queries like the distinct query in our example. While leading to a significant performance benefit if matched by a user query, the major drawback of materialized views is their ability to handle updates. Typically, they need to be re-computed when updates occur to keep them consistent with the actual database. Alternatively, these expensive recomputations can be delayed and run regularly, if minor inconsistencies are acceptable for the user application. As materialized views are not offered by Actian Vector, we simulate this approach by storing the materialized information in a separate table and manually rewrtiting queries. For sort queries, we compared PatchIndexes against SortKeys, which physically sorts data on the given SortKey column. This way, sort queries can be translated to simple scan queries. As a drawback, physically reordering data is a very costly operation and maintaining this order in case of updates requires additional effort. Last, we evaluated join queries by comparing the PatchIndex approach against JoinIndexes, which materialize foreign key joins as an additional table column. If a SortKey is defined on the table holding the primary key of the join, the foreign key related table is ordered similarly, so that a MergeJoin becomes possible to join both tables.

\vspace{-0.25em}
\subsection{Sharded bitmap microbenchmarks}
\begin{figure}
\begin{tikzpicture}
\pgfplotsset{
    width=8cm,
    height=4.5cm,
    compat=1.3,
    legend style={at={(-0.1,1.2)},anchor=west},    
    legend columns=3, 
}

\begin{axis}[
  axis y line*=left,
  ymin=0, ymax=5,
  ytick={0,1,2,3,4,5},
  xlabel=Shard size in bits (log),
  xtick={0,1,2,3,4,5,6,7,8,9, 10, 11},
  xticklabels={$2^{8}$, $2^{9}$, $2^{10}$, $2^{11}$, $2^{12}$, $2^{13}$, $2^{14}$, $2^{15}$,$2^{16}$, $2^{17}$, $2^{18}$, $2^{19}$},
  ylabel=Bulk delete runtime {[s]},
]
\addplot[smooth,mark=x,red]
  coordinates{
  (0, 0.8)(1, 0.61)(2, 0.54)(3, 0.48)(4, 0.36)(5, 0.21)(6, 0.23)(7, 0.33)(8, 0.58)(9,1.11)(10, 2.27)(11,4.47)}; \label{plot_one}
,
\addplot[smooth,mark=triangle,green!70!black]
  coordinates{
  (0, 0.84)(1, 0.63)(2, 0.53)(3, 0.46)(4, 0.35)(5, 0.23)(6, 0.18)(7, 0.27)(8, 0.52)(9,0.8)(10, 1.61)(11,3.25)}; \label{plot_two}
\end{axis}

\begin{axis}[
  axis y line*=right,
  axis x line=none,
  ymin=0, ymax=30,
  ytick={0,5,10,15,20,25,30},
  xtick={0,1,2,3,4,5,6,7,8,9, 10, 11},
  xticklabels={$2^{8}$, $2^{9}$, $2^{10}$, $2^{11}$, $2^{12}$, $2^{13}$, $2^{14}$, $2^{15}$,$2^{16}$, $2^{17}$, $2^{18}$, $2^{19}$},
  ylabel=Memory overhead {[\%]},
]
\addlegendimage{/pgfplots/refstyle=plot_one}\addlegendentry{Parallel}
\addlegendimage{/pgfplots/refstyle=plot_two}\addlegendentry{Parallel \& vect.}
\addplot[smooth,mark=*,blue]
  coordinates{(0,25)(1, 12.5)(2, 6.25)(3, 3.125)(4, 1.56)(5, 0.78)(6, 0.39)(7, 0.2)(8,0.1)(9,0.05)(10, 0.025)(11, 0.0125)
}; \addlegendentry{Mem. overhead}
\end{axis}
\end{tikzpicture}
\vspace{-1em}
\caption{Sharded bitmap bulk delete runtime for 1M elements and memory overhead of sharding depending on shard size}
\vspace{-1em}
\label{shard_size_impact}
\end{figure}
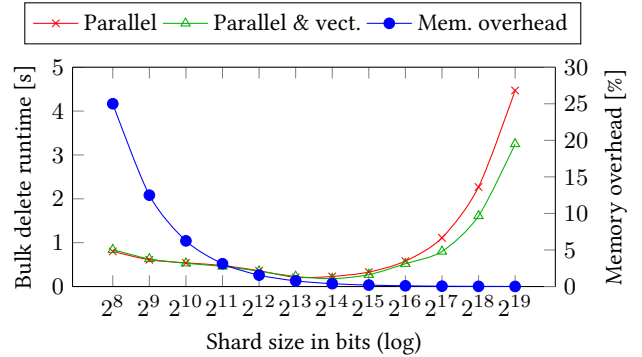

\vspace{-0.25em}
Choosing the shard size is crucial for the performance and the memory overhead of the sharded bitmap data structure. A small shard size leads to a large memory overhead due to the additional start values and deletion performance overhead for their adaption. On the contrary, a large shard size leads to a negligible memory overhead, but results in shifting large amounts of memory, which was the problem we intended to avoid with the design of the sharded bitmap. The runtime of bulk deleting 1M elements (randomly chosen but fixed) from a sharded bitmap of size 100M is shown in Figure~\ref{shard_size_impact} for the parallel and parallel~\&~vectorized bulk delete implementation. First, we can locate a clear minimum runtime at a shard size of $2^{14}$ bits. Below this minimum, the overhead of preprocessing and thread starting is not worth the benefit of multithreading and above this minimum the shifting effort starts to dominate the runtime again. Second, we can observe that vectorization impacts performance only in a minor way at these relatively small shard sizes, but gets more impactful the larger the shard size is chosen. Regarding memory consumption, a 64~Bit start value is stored for each shard, resulting in a memory overhead of $\frac{64}{shard\_size} \cdot 100\%$. Choosing the shard size of $2^{14}$ bits therefore leads to a memory overhead of 0.39\% for sharding.

\begin{table}
\centering
\begin{tabular}{lcc}
 \hline
  & \textbf{Bitmap} & \textbf{Sharded bitmap} \\
 \hline
 \textbf{Sequential Set} & $8.1\;ns$ & $16\;n$s\\

 \textbf{Sequential Get} &$5.1\;ns$ & $10.1\;ns$\\

 \textbf{Seq. Delete} & $3.1 \cdot 10^6\; ns$ & $3400\;ns$ \\

 \textbf{Seq. Bulk Delete} & $-$  & $180\;ns$ \\
  \hline
\end{tabular}

\caption{Bitmap operator runtimes per element for bitmap with 100M elements and shard size $2^{14}$ bits}
\vspace{-2em}
\label{bitmap_operators}
\end{table}

Table~\ref{bitmap_operators} shows the latency per element of the bitmap operators that are relevant for the PatchIndex. Comparing an ordinary bitmap with the sharded bitmap approach, the virtual sharding leads to a small overhead for bit access operations, which is caused by determining the group an element \mbox{belongs to}. In terms of delete operations, the sharded bitmap performs three orders of magnitude faster than the ordinary bitmap, which improves to another order of magnitude when using the bulk delete operation to delete 1M elements. For the ordinary bitmap structure the delete runtime is size dependent and linearly increases/decreases when changing the bitmap size. Consequently, the sharded bitmap data structure significantly outperforms the ordinary bitmap in terms of deletion support, which particularly holds for large bitmap sizes, and justifies the small bit access overhead and small memory overhead for use cases where update support is important. 

\vspace{-0.25em}
\subsection{PatchIndex microbenchmarks}\label{microbench}
\vspace{-0.25em}
For our microbenchmarks, we designed a data generator \cite{data_gen_github} that varies the exception rates to given constraints. The data consists of 1B tuples with two columns, a unique \emph{key} column and a \emph{value} column that shows the desired data distribution. With the tuple width of 128\;Bytes this results in a dataset size of 128\;GB. In order to exploit parallel data processing, we partition the datasets on the \emph{key} column into 24 partitions. As the \emph{key} column is unique, this results in partitions of nearly equal size. For the uniqueness constraint, exceptions of the \emph{value} column are equally distributed into 100K values, while the remaining values are unique and differ from the values of the exceptions. For the sorting constraint, exceptions are randomly chosen and all remaining values form a sorted sequence in ascending order. For both constraints, exceptions are randomly placed in the datasets. As the datasets are generated once, the randomness does not impact the comparability of the evaluation results. For the evaluation of update operations, we chose the dataset with exception rate $e = 0.5$. This choice has no impact on update query performance, as bits of updated patches have to be accessed independently from being patches before the update.

The materialization is realized in different ways for the constraints in the microbenchmarks. For the uniqueness constraint, we materialized a table containing all unique values of the \emph{value} column using a distinct query. As a result, a distinct query on the \emph{value} column is replaced by a scan query on the materialized view without the need for an expensive aggregation. For the sorting constraint, we materialized the order information using a SortKey on the \emph{value} column. This way, the data of the real table is physically re-ordered according to the \emph{value} column. Queries that include a sort operator therefore just scan the table. As the table is partitioned, an additional merge step of the tuples from each partition is necessary to preserve the global order of the query results.

\subsubsection{Performance impact}
With this experiment, we want to prove the query performance improvement for different exception rates when using PatchIndexes. We run a distinct query and a sort query respectively while varying the exception rates for the uniqueness and sorting constraint in the dataset. 

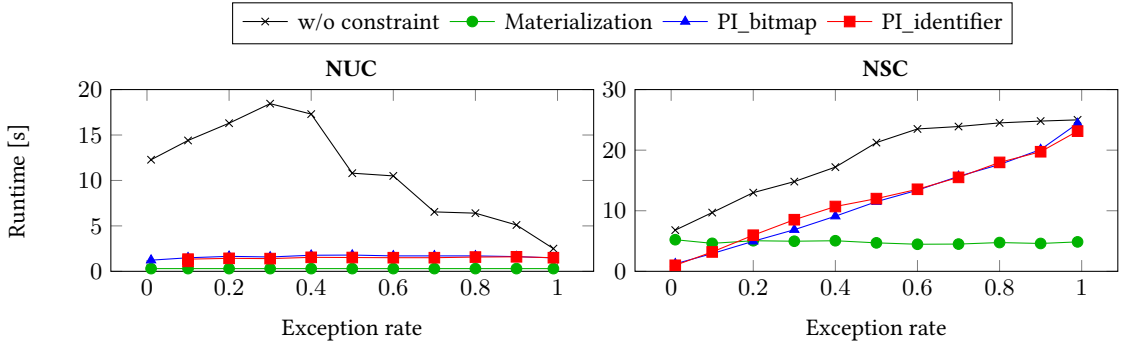
\begin{figure*}
\centering
\begin{tikzpicture}

    \begin{axis}[%
    name=unique,
    height=4cm,
    width=8cm,
    ylabel style={align=center},
    ylabel=Runtime {[s]},
    xlabel=Exception rate,
    ymin=0, ymax=20]
    \addplot[mark=x,black]
  coordinates{
    (0.01,12.28)(0.10,14.4)(0.20,16.3)(0.30,18.45)(0.40,17.3)(0.50,10.8)(0.60,10.5)(0.70,6.54)(0.80,6.4)(0.90,5.1)(0.99,2.5)
};
\label{pgfplots:ref}
\addplot[mark=*,green!70!black]
  coordinates{
    (0.01,0.3)(0.10,0.3)(0.20,0.3)(0.30,0.3)(0.40,0.3)(0.50,0.3)(0.60,0.3)(0.70,0.3)(0.80,0.3)(0.90,0.3)(0.99,0.3)
};
\label{pgfplots:mat}
\addplot[mark=triangle*,blue]
  coordinates{
       (0.01,1.23)(0.10,1.49)(0.20,1.65)(0.30,1.58)(0.40,1.77)(0.50,1.79)(0.60,1.7)(0.70,1.7)(0.80,1.7)(0.90,1.62)(0.99,1.48)
};
\label{pgfplots:pi_bitmap}
\addplot[mark=square*,red]
  coordinates{
       (0.1,1.1)(0.10,1.34)(0.20,1.43)(0.30,1.4)(0.40,1.54)(0.50,1.52)(0.60,1.5)(0.70,1.5)(0.80,1.57)(0.90,1.6)(0.99,1.5)
};
\label{pgfplots:pi_rid}
    \end{axis}

    \begin{axis}[%
    name=sort,
    height=4cm,
	width=8cm,
    xlabel=Exception rate,
    at=(unique.right of south east), anchor=left of south west,
    ymin=0, ymax=30]
    \addplot[mark=x,black]
  coordinates{
      (0.01,6.8)(0.10,9.7)(0.20,12.99)(0.30,14.81)(0.40,17.2)(0.50,21.25)(0.60,23.5)(0.70,23.9)(0.80,24.5)(0.90,24.8)(0.99,25)
};
\addplot[mark=*,green!70!black]
  coordinates{
     (0.01,5.22)(0.10,4.62)(0.20,5.05)(0.30,4.97)(0.40,5.05)(0.50,4.7)(0.60,4.47)(0.70,4.5)(0.80,4.75)(0.90,4.6)(0.99,4.86)
};
\addplot[mark=triangle*,blue]
  coordinates{
    (0.01,1.24)(0.10,2.93)(0.20,4.96)(0.30,6.87)(0.40,9.1)(0.50,11.5)(0.60,13.39)(0.70,15.68)(0.80,17.63)(0.90,20.1)(0.99,24.4)
};
\addplot[mark=square*,red]
  coordinates{
      (0.01,1)(0.10,3.2)(0.20,5.96)(0.30,8.52)(0.40,10.7)(0.50,12)(0.60,13.544)(0.70,15.51)(0.80,17.98)(0.90,19.74)(0.99,23.14)
};    
    
    \end{axis}
    
    \matrix[
    matrix of nodes,
    draw,
    inner sep=0.2em,
    draw
  ]
  at(6.8,3.3){
    \ref{pgfplots:ref}& w/o constraint &
    \ref{pgfplots:mat}& Materialization &
    \ref{pgfplots:pi_bitmap}& PI\_bitmap &
    \ref{pgfplots:pi_rid}& PI\_identifier\\};
    
   \node[] at (3.2,2.7) {\textbf{NUC}};
   \node[] at (10.3,2.7) {\textbf{NSC}};
\end{tikzpicture}
\vspace{-1em}
\caption{Runtimes of a distinct/sort query with varying exception rate}

\label{perf_runtimes}
\end{figure*}

\begin{figure*}
\centering
\begin{tikzpicture}

    \begin{axis}[%
    name=unique,
    height=4cm,
    width=8cm,
    ylabel style={align=center},
    ylabel=Runtime {[s]},
    xlabel=Exception rate,
    ymin=0, ymax=40]

\addplot[mark=*,green!70!black]
  coordinates{
    (0.01,13.7)(0.10,15.7)(0.20,18.5)(0.30,22.8)(0.40,20.3)(0.50,14.78)(0.60,15.2)(0.70,8.2)(0.80,8.1)(0.90,8.9)(0.99,2.6)
};
\label{pgfplots:mat}
\addplot[mark=triangle*,blue]
  coordinates{
       (0.01,18.5)(0.10,24.1)(0.20,25.2)(0.30,30.2)(0.40,28.7)(0.50,19.6)(0.60,19.6)(0.70,12.85)(0.80,13.5)(0.90,10.2)(0.99,5.7)
};
\label{pgfplots:pi_bitmap}
\addplot[mark=square*,red]
  coordinates{
       (0.01,18.3)(0.10,26.5)(0.20,31.4)(0.30,36.1)(0.40,34)(0.50,25.7)(0.60,24.9)(0.70,14.96)(0.80,13.7)(0.90,14.1)(0.99,5.7)
};
\label{pgfplots:pi_rid}
    \end{axis}

    \begin{axis}[%
    name=sort,
    height=4cm,
	width=8cm,
    xlabel=Exception rate,
    at=(unique.right of south east), anchor=left of south west,
    ymin=0, ymax=120]

\addplot[mark=*,green!70!black]
  coordinates{
     (0.01,15)(0.10,23.5)(0.20,35.9)(0.30,63.5)(0.40,73.5)(0.50,95.6)(0.60,109)(0.70,113)(0.80,105)(0.90,90)(0.99,43.62)
};
\addplot[mark=triangle*,blue]
  coordinates{
    (0.01,10.23)(0.10,17.53)(0.20,21.55)(0.30,25.16)(0.40,28.54)(0.50,31.48)(0.60,34.19)(0.70,36.97)(0.80,50.37)(0.90,45.15)(0.99,31.59)
};
\addplot[mark=square*,red]
  coordinates{
      (0.01,16.81)(0.10,22.96)(0.20,29.9)(0.30,33.19)(0.40,34.14)(0.50,43.5)(0.60,45.57)(0.70,48.21)(0.80,52.15)(0.90,49.8)(0.99,34.04)
};    
    
    \end{axis}
    
%
   \node[] at (3.2,2.7) {\textbf{NUC}};
   \node[] at (10.3,2.7) {\textbf{NSC}};
\end{tikzpicture}
\vspace{-1em}
\caption{Runtime for materialization/index creation for varying exception rate}

\label{creation_runtimes}
\end{figure*}
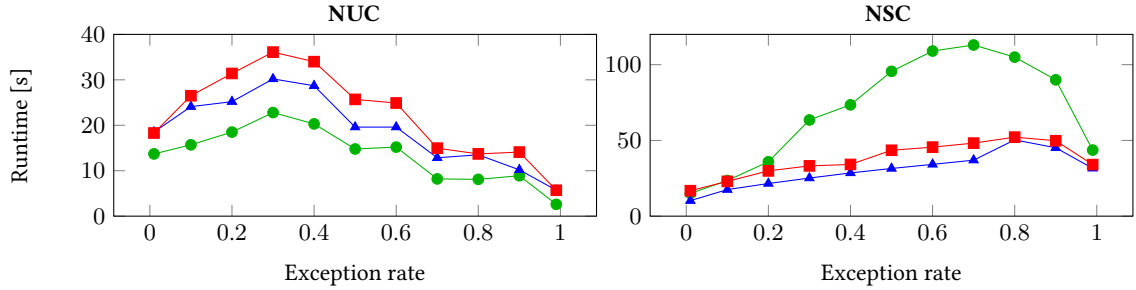

\begin{figure*}
\centering
\begin{tikzpicture}

    \begin{axis}[%
    name=ins_unique,
    height=3.5cm,
    width=5.5cm,
    xticklabels={5,10,50,100,500,1000},
    ylabel style={align=center},
    ylabel=\textbf{NUC}\\ Runtime {[s]},
	xtick={1,...,6},
    ymin=0, ymax=200]
    \addplot[mark=x,black]
  coordinates{
    (1,6.05)(2,3.2)(3,1.03)(4,0.62)(5,0.3)(6,0.36)
};
\label{pgfplots:ref}
\addplot[mark=*,green!70!black]
  coordinates{
    (3,668)(4,276)(5,54)(6,28)
};
\label{pgfplots:mat}
\addplot[mark=triangle*,blue]
  coordinates{
    (1,111)(2,57.5)(3,11.8)(4,6.09)(5,1.5)(6,1)
};
\label{pgfplots:pi_bitmap}
\addplot[mark=square*,red]
  coordinates{
    (2,663)(3,134)(4,69.7)(5,15.2)(6,8.5)
};
\label{pgfplots:pi_rid}
    \end{axis}

    \begin{axis}[%
    name=mod_unique,
    height=3.5cm,
    width=5.5cm,
    xticklabels={5,10,50,100,500,1000},
	xtick={1,...,6},
    at=(ins_unique.right of south east), anchor=left of south west,
    ymin=0, ymax=200]
    \addplot[mark=x,black]
  coordinates{
    (1,22.8)(2,11.8)(3,2.6)(4,1.3)(5,0.38)(6,0.26)
};
\addplot[mark=*,green!70!black]
  coordinates{
    (3,400)(4,194)(5,36.7)(6,18.4)
};
\addplot[mark=triangle*,blue]
  coordinates{
    (1,97.4)(2,49.1)(3,10.6)(4,5.1)(5,1)(6,0.5)
};
\addplot[mark=square*,red]
  coordinates{
    (2,430)(3,87.9)(4,46.5)(5,10.4)(6,6.6)
};    
    
    \end{axis}

 \begin{axis}[%
    name=del_unique,
    height=3.5cm,
    width=5.5cm,    
    xticklabels={5,10,50,100,500,1000},
	xtick={1,...,6},
    at=(mod_unique.right of south east), anchor=left of south west,
    ymin=0, ymax=200]
    \addplot[mark=x,black]
  coordinates{
    (1,15.4)(2,7.03)(3,1.83)(4,1)(5,0.3)(6,0.23)
};
\addplot[mark=*,green!70!black]
  coordinates{
    (3,391)(4,195)(5,36.8)(6,18.23)
};
\addplot[mark=triangle*,blue]
  coordinates{
    (1,18.2)(2,9.4)(3,2.5)(4,1.23)(5,0.48)(6,0.35)
};
\addplot[mark=square*,red]
  coordinates{
    (1,37.2)(2,20)(3,4.58)(4,2.3)(5,0.53)(6, 0.33)
};    
    
    \end{axis}

    \begin{axis}[%
    name=ins_sort,
    height=3.5cm,
    width=5.5cm,
    ylabel style={align=center},
    ylabel=\textbf{NSC} \\ Runtime {[s]},
    xticklabels={5,10,50,100,500,1000},
	xtick={1,...,6},
     at=(ins_unique.below south west), anchor=above north west,
    ymin=0, ymax=200]
    \addplot[mark=x,black]
  coordinates{
    (1,9.2)(2,4.2)(3,1.12)(4,0.56)(5,0.35)(6,0.33)
};
\addplot[mark=*,green!70!black]
  coordinates{
    (4,398)(5,76)(6,37.7)
};
\addplot[mark=triangle*,blue]
  coordinates{
    (1,56.3)(2,28.6)(3,6.1)(4,3.04)(5,0.82)(6,0.57)
};
\addplot[mark=square*,red]
  coordinates{
    (2,451)(3,92.5)(4,48.4)(5,9.37)(6, 4.7)
};
    \end{axis}

    \begin{axis}[%
    name=mod_sort,
    height=3.5cm,
    width=5.5cm,
    xticklabels={5,10,50,100,500,1000},
	xtick={1,...,6},
	xlabel=\textbf{Update granularity},
    at=(ins_sort.right of south east), anchor=left of south west,
    ymin=0, ymax=200]
    \addplot[mark=x,black]
  coordinates{
    (1,20.7)(2,10.4)(3,2.3)(4,1.3)(5,0.28)(6,0.25)
};
\addplot[mark=*,green!70!black]
  coordinates{
    (4,413)(5,81.8)(6,41.5)
};
\addplot[mark=triangle*,blue]
  coordinates{
    (1,20.6)(2,11.3)(3,2.2)(4,1.3)(5,0.33)(6,0.24)
};
\addplot[mark=square*,red]
  coordinates{
    (1,26.8)(2,15.1)(3,5)(4,3.3)(5,2)(6, 1.8)
};    
    
    \end{axis}

 \begin{axis}[%
    name=del_sort,
    height=3.5cm,
    width=5.5cm,
    xticklabels={5,10,50,100,500,1000},
	xtick={1,...,6},
    at=(mod_sort.right of south east), anchor=left of south west,
    ymin=0, ymax=200]
    \addplot[mark=x,black]
  coordinates{
    (1,15.8)(2,7.9)(3,2)(4,1)(5,0.3)(6,0.23)
};
\addplot[mark=*,green!70!black]
  coordinates{
    (4,355)(5,69.5)(6,35.9)
};
\addplot[mark=triangle*,blue]
  coordinates{
    (1,19)(2,10)(3,2.56)(4,1.32)(5,0.39)(6,0.33)
};
\addplot[mark=square*,red]
  coordinates{
    (1,36.8)(2,21.4)(3,4.57)(4,2.47)(5,0.55)(6, 0.31)
};    
    
    \end{axis}
    
    
    \node[] at (2,2.3) {\textbf{INSERT}};
   \node[] at (6.7,2.3) {\textbf{MODIFY}};
   \node[] at (11.4,2.3) {\textbf{DELETE}};
\end{tikzpicture}
\vspace{-1em}
\caption{Update performance for inserting/updating/deleting 1000 tuples and varying update granularities}

\label{update_runtimes}
\end{figure*}
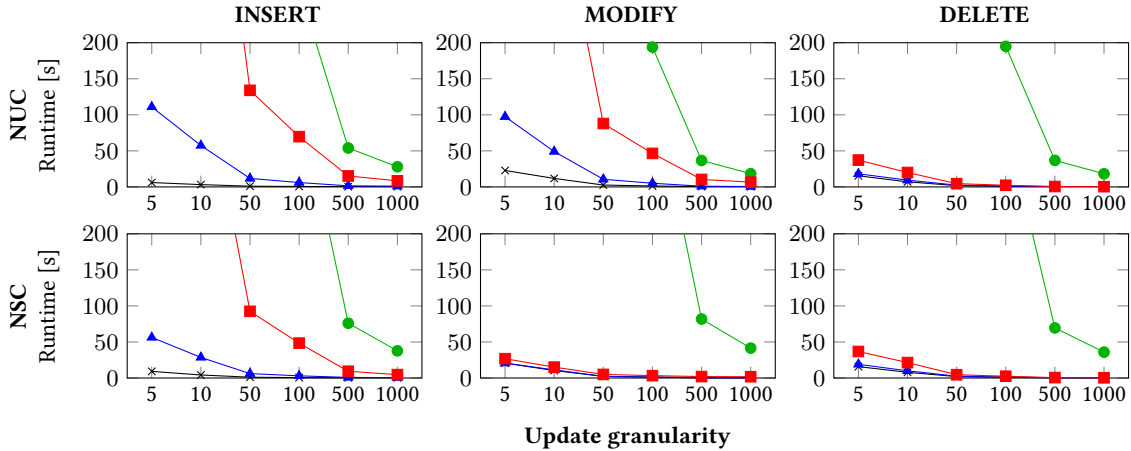

Figure~\ref{perf_runtimes} shows the results of this experiment. For the uniqueness constraint, reference runtimes without any constraint definition increase with increasing exception rates, before decreasing starting from an exception rate of 0.3. With increasing exception rates, the number of distinct tuples and therefore the number of aggregation groups decrease. The runtime behaviour is then caused by the inference of a reduced hash table size and the increased communication costs, as the system uses a shared hash table build plan for the aggregation. The use of a materialized view shows a nearly constant runtime as the query only scans the materialized result. Using a PatchIndex shows a significant performance benefit compared to the reference runtimes with performance comparable to the materialized view. Both PatchIndex design approaches perform similarly and show a slight increase in runtime with increasing exception rates, caused by more tuples being processed in the aggregation. As a result, the actual performance gain compared to the reference runtime shrinks with increasing exception rates, but using a PatchIndex does not impact runtimes in a negative way for the evaluated cases.

For the sorting constraint, reference runtimes increase with increasing exception rates, which is caused by the pivoting strategy of the internal QuickSort implementation, behaving better the more sorted the input sequence already is. Using a SortKey as materialization shows a constant runtime, although slightly slower than the scan query for the uniqueness constraint, as merging the sorted partitions is necessary. Additionally, the query still performs a sort operator to ensure the sorting, resulting in a slightly worse performance than using a PatchIndex for small exception rates. Using the PatchIndex shows a significant performance gain compared to the reference runtimes. Runtimes increase as expected with increasing exception rates as more tuples have to be processed by the sort operator, so the relative performance gain shrinks with increasing exception rate. Again, using a PatchIndex does not impact runtimes in a negative way for the evaluated cases.

\subsubsection{Memory consumption} The memory consumption of a PatchIndex is independent from the materialized constraint and shown in Table~\ref{mem_consumption}. While the bitmap-based approach has a constant total memory consumption (1~bit per tuple + sharded bitmap overhead), the memory consumption of the identifier-based approach grows linearly with the number of exceptions. Hence, the bitmap-based approach has a lower memory consumption for cases with exception rate $e > 0.0158$. In contrast, the materialized view for the uniqueness constraint materializes every unique value (100K unique values + exceptions), leading to a significantly higher memory consumption than the PatchIndex for most cases.

\begin{table}
\centering
\vspace{0.5em}
\resizebox{\columnwidth}{!}{%
\begin{tabular}{lccc}
 \hline
  & \textbf{PI\_bitmap} & \textbf{PI\_identifier} & \textbf{Mat. view (NUC)} \\
 \hline
 General & $ t/8 \cdot 1.0039 \;B$ & $e \cdot t \cdot 8 \;B$& $(10^5 + (1-e) \cdot t) \cdot 8 \;B$ \\

 \textbf{$e=0.01$} & $125.48\;\mathit{MB}$ & $80\;\mathit{MB}$&  $7.9\;\mathit{GB}$\\

   \textbf{$e=0.2$} & $125.48\;\mathit{MB}$ & $1.6\;\mathit{GB}$&  $6.4\;\mathit{GB}$\\
  \hline
\end{tabular}
}
\caption{Memory consumption for example dataset of 128\;GB / $t = 10^9$ tuples}
\vspace{-3em}
\label{mem_consumption}
\end{table}

\subsubsection{Creation time}

Another important fact for the usability of data structures is their creation effort. Both PatchIndex and materialization are intended to be used multiple times after being created. Otherwise the creation effort would not be worth compared to the achieved query speedups. Figure~\ref{creation_runtimes} shows the comparison of the runtimes for creating the PatchIndex and the materialization for both constraints. For the uniqueness constraint, the runtimes follow the reference runtimes of Figure~\ref{perf_runtimes}, as the distinct query is pre-computed and materialized. The effort to create a PatchIndex is slightly higher than the materialization, as the information about exceptions have to be filled into the index structure. Furthermore, the bitmap-based approach performs better than the identifier-based approach, as here only bits have to be set in an already allocated bitmap instead of maintaining a growing list of identifiers.

For the sorting constraint, creating the SortKey takes a huge amount of time, as this physically reorders the table data. In comparison, creating a PatchIndex is more efficient. The increasing exception rate leads to an increasing number of comparisons in the longest sorted subsequence algorithm, while decreasing the length of the sorted sequence and therefore decreasing the effort to reconstruct it. The inference of both parts leads to the observed runtimes. Next to the worse creation performance of a SortKey, it's definition is also limited to one per table, as it physically reorders the data. In comparison, PatchIndexes can be defined multiple times per table on different columns, as it does not change the way data is physically stored. Again, the bitmap-based approach performs better than the identifier-based approach.

\subsubsection{Insert}
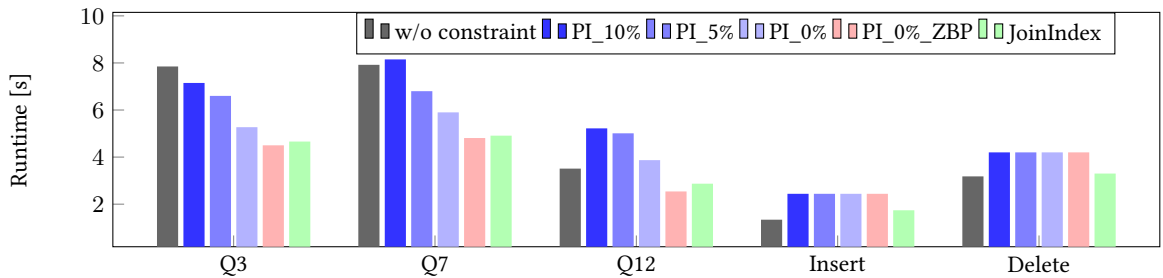
\begin{figure*}
\centering
\begin{tikzpicture}
\begin{axis}[
    ybar,tick pos=left,
    width=15.5cm,
    height=4.7cm,
    enlargelimits=0.15,
    legend style={at={(0.595,1)},
      anchor=north,legend columns=-1},
    ylabel={Runtime {[s]}},
    symbolic x coords={Q3,Q7,Q12, Insert, Delete},
    xtick=data,
    xtick align=inside,
    nodes near coords align={vertical},
    bar width=8,
    ymax=9
    ]
\addplot [draw=none, fill=black!60] coordinates {(Q3,7.85) (Q7,7.92) (Q12,3.51) (Insert,1.34)(Delete,3.18)};
\addplot [draw=none, fill=blue!80] coordinates {(Q3,7.15) (Q7,8.15) (Q12,5.22)(Insert,2.44)(Delete,4.20)};
\addplot [draw=none, fill=blue!50] coordinates {(Q3,6.6) (Q7,6.8) (Q12,5.01) (Insert,2.44)(Delete,4.20)};
\addplot [draw=none, fill=blue!30]coordinates {(Q3,5.27) (Q7,5.9) (Q12,3.87)(Insert,2.44)(Delete,4.20)};
\addplot [draw=none, fill=red!30]coordinates {(Q3,4.5) (Q7,4.81) (Q12,2.54)(Insert,2.44)(Delete,4.20)};
\addplot [draw=none, fill=green!30] coordinates {(Q3,4.66) (Q7,4.91) (Q12,2.87)(Insert,1.74)(Delete,3.3)};
\legend{w/o constraint,PI\_10\%, PI\_5\%, PI\_0\%, PI\_0\%\_ZBP,JoinIndex}
\end{axis}
\end{tikzpicture}
\vspace{-1em}
\caption{TPC-H query performance}
\vspace{-1em}
\label{tpch_runtimes}
\end{figure*}

Based on the dataset with an exception rate $e = 0.5$, we inserted 1000 tuples per run into the database and varied the granularity of the insert operations between 5 tuples per operation (200 queries in total) and 1000 tuples per operation (1 query in total) to capture the impact of trickle and bulk inserts. The left column of Figure~\ref{update_runtimes} shows the total runtimes to insert the 1000 tuples. First, we can observe that recomputing or maintaining the materialization for each insert operation produces a tremendous overhead to the reference runtime without any constraint definition, making it especially not usable for trickle updates. In contrast, PatchIndexes support insert handling in a more efficient way. For both constraints, the identifier-based design approach performs worse than the bitmap-based approach, caused by keeping the list of identifiers sorted. For the uniqueness constraint, every insert operation invokes the insert handling query shown in Figure~\ref{qet_insert_handling}, resulting in an overhead for fine-grained inserts. Similarly, the sorting constraint invokes the execution of the longest sorted subsequence algorithm on the inserted tuples for every insert query. Nevertheless, the added overhead is smaller compared to the uniqueness constraint.
For insert granularities of 50 tuples per operation or higher, the overhead is negligible for both constraints and defining a PatchIndex on a column does not impact insert runtimes in a remarkable way. Comparing the plot for the bitmap-based approach and for the materialization, this result also leads to another consequence. Keeping the runtime nearly equal by fixing a value on the \mbox{y-axis}, update cycles can be chosen 50 times more frequently for the uniqueness constraint and 100 times more frequently for the sorting constraint when using a PatchIndex instead a materialization. This increases the ability of the system to keep the materialized information consistent with the actual dataset.

\subsubsection{Modify}
Similar to the insert experiment, we here updated 1000 tuples of the dataset with exception rate $e=0.5$ for different granularities. For the uniqueness constraint, we can observe a similar behaviour than for the insert support, as both perform the same query to handle the updates and keep the PatchIndex accurate. For the sorting constraint, updates must be included to the exceptions in the PatchIndex, as they may destroy the sorted subsequence . This can be done efficiently without the need for an additional query, leading to nearly no overhead compared to the reference runtime.  

\subsubsection{Delete}
The right column of Figure~\ref{update_runtimes} shows the total runtime for deleting 1000 tuples from the dataset with exception rate $e=0.5$ with varying granularities. As the PatchIndex just drops information about deleted tuples, handling deletes is a very efficient operation and adds nearly no overhead to the operation runtime. Furthermore, we can observe that the identifier-based approach performs worse than the bitmap-based approach, caused by the fact that identifiers have to be decreased when a tuple with a lower identifier is deleted, while handling deletes in the bitmap-based approach is realized using efficient bulk delete operations of the sharded bitmap.

\vspace{-0.25em}
\subsection{TPC-H}
\vspace{-0.25em}

\begin{figure*}
\centering
\subfloat[PatchIndex]{
\begin{tikzpicture}%
\clip (-1.9,-1.9) rectangle (1.9, 1.9);
\tkzKiviatDiagram[label space=.75, scale=0.6, lattice = 4]{C, M, P, U}%
\tkzKiviatLine[color=red, fill=red!20](2.5,2.5,2.5,3)%
\end{tikzpicture}%
}%
\subfloat[Mat. View]{
\begin{tikzpicture}%
\clip (-1.9,-1.9) rectangle (1.9, 1.9);
\tkzKiviatDiagram[label space=.75,scale=0.6, lattice = 4]{C, M, P, U}%
\tkzKiviatLine[color=red, fill=red!20](2.5,1,3.5,1)%

\end{tikzpicture}%
}%
\subfloat[SortKey]{
\begin{tikzpicture}%
\clip (-1.9,-1.9) rectangle (1.9, 1.9);
\tkzKiviatDiagram[label space=.75,scale=0.6, lattice = 4]{C, M, P, U}%
\tkzKiviatLine[color=red, fill=red!20](1.5,3.5,3,1.5)%

\end{tikzpicture}%
}%
\subfloat[JoinIndex]{
\begin{tikzpicture}%
\clip (-1.9,-1.9) rectangle (1.9, 1.9);
\tkzKiviatDiagram[label space=.75,scale=0.6, lattice = 4]{C, M, P, U}%
\tkzKiviatLine[color=red, fill=red!20](1,2,3,3)%
\end{tikzpicture}%
}%
\vspace{-0.5em}
\caption{Qualitative comparison of PatchIndex against evaluated approaches in terms of Creation effort~(C), Memory/Storage overhead~(M), Performance impact~(P) and Updatability~(U). (Higher score means ``better'')}
\vspace{-1.5em}
\label{radarplots}
\end{figure*}
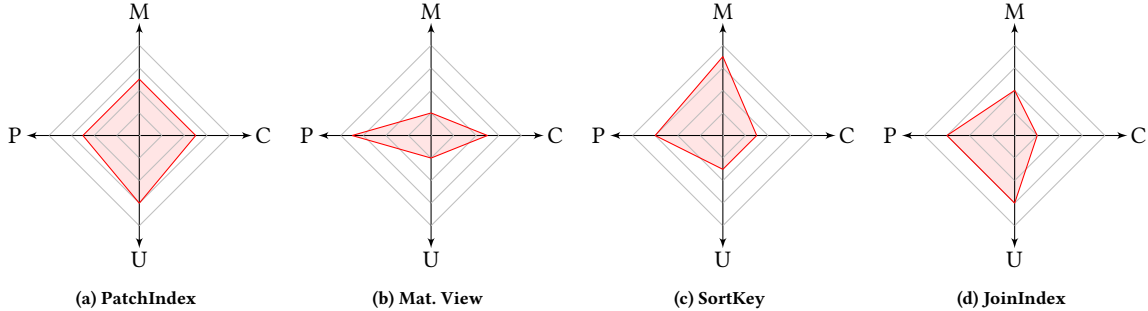


The TPC-H benchmark \cite{boncz_tpc-h_2014} is a well-known and well-understood benchmark used for performance evaluation of analytical DBMS for many years. Besides analytical queries, the benchmark also contains update sets for table inserts and table deletes that can be used to evaluate update support. In our experiments, we used the benchmark at scale factor SF\;1000.

We decided to focus on the largest join in the benchmark, which is the join between the lineitem and orders table. Although the benchmark only contains clean data with perfect constraints, we manually manipulated the data order of the lineitem table in order to introduce exceptions to the sorting constraint, resulting in three datasets with 0\%, 5\% and 10\% exceptions. While we stored the orders table in a sorted way, we evaluated the impact of a PatchIndex and a JoinIndex defined on the lineitem table for query performance and chose a subset of three queries that includes this join. We used the bitmap-based PatchIndex approach in our experiments, which showed in Section~\ref{microbench} to perform better than the identifier-based approach. Besides different exception rates, we evaluated two options for the dataset without exceptions. The first option uses the PatchIndex optimizations described in Section~\ref{back_queryproc} and is therefore the general case that is comparable against the reference runtime in a fair way. Nevertheless, these queries contain unnecessary overhead as the constraints in the benchmark do not contain any exceptions. In the second option we therefore enable zero-branch-pruning (ZBP), which is a technique of removing subtrees of a query plan that are ensured to not produce any results. During query optimization query plans are annotated with cardinality estimations. If these estimates are ensured to be zero, e.g. by evaluating integrity constraints or indexes, the query rewriter can drop the respective subtrees from the query plan. This way, the subtree that would process the patches is pruned from the query plan, resulting in better performance as it drops all overhead introduced by cloning the query subtrees. Although not being the general use case, this option can therefore be used to compare results against the JoinIndex runtimes.

Figure~\ref{tpch_runtimes} shows the results of the experiment for queries as well as for update sets. In general, the exception rate has no impact on reference runtimes without constraint, JoinIndex runtimes or update runtimes with PatchIndexes. We can observe that the PatchIndex benefit on join queries depends on the size of the join and the exception rate. For Q3, which contains the largest join, query runtime decreases with decreasing exception rate and impacts query performance in a positive way even for an exception rate of 10\%. The same holds for Q7, which however shows worse performance compared to the reference runtime for 10\% exceptions. As the PatchIndex impact grows with lower exception rates, PatchIndex runtimes nearly reach the JoinIndex runtimes for an exception rate of 0\%. Additionally activating zero-branch-pruning reduces the overhead introduced by the Patchindex optimization, leading the queries to run 43\% and 40\% faster compared to the reference runtime. With zero-branch-puning, runtimes are also slightly faster than using the JoinIndex, which is a full materialization of the join. This is caused by a small additional scan effort in the JoinIndex query, as here the index is materialized in an additional table column.
Query Q12 shows a different behaviour, as this already short-running query is impacted in a negative way when using a PatchIndex. In opposite to queries Q3 and Q7, the join in this query is very small due to prior selections. Therefore, the added overhead of cloning subtrees is larger than the gained benefit of exchanging the HashJoin with a MergeJoin in the PatchIndex optimization. Nevertheless, enabling zero-branch-pruning leads to a performance benefit of 28\%, which is slightly faster than the JoinIndex query again. 
In terms of update support, zero-branch-pruning does not have any impact on performance. For the insert set (inserting 0.5M tuples) and the delete set (deleting 6M tuples) we can observe a slight performance overhead for both materialization approaches. Here the JoinIndex performs slightly better than the PatchIndex, as updates are handled in-memory by the Positional Delta-Tree \cite{heman_positional_2010} structure. The creation effort for both approaches is excluded from Figure~\ref{tpch_runtimes} for scaling purpose. Here, creating a PatchIndex takes 100 seconds, which is significantly faster than creating a JoinIndex, which takes around 600 seconds.

In contrary to distinct and sort queries, this experiments shows that using PatchIndexes in join queries is a trade-off, as the overhead of cloned subtrees might negate the benefit of using a MergeJoin for most tuples. As described in Section~\ref{cost_model}, these plans would not be chosen by the optimizer.
Although the benchmark (theoretically) only contains perfect constraints, it is an example for another advantage of the PatchIndex approach. Even if a dataset is clean at a point in time, it may become unclean in the future by update operations. While these updates would be aborted with the definition of usual constraints, PatchIndexes would allow the updates and the respective transition from a perfect constraint to an approximate constraint.

\vspace{-0.25em}
\subsection{Resume}
\vspace{-0.25em}
In our evaluation, we compared the generic PatchIndex approach against materialized views, SortKeys and JoinIndexes as specialized materialization approaches for different queries. Figure~\ref{radarplots} shows a qualitative comparison of the discussed approaches, proving that the PatchIndex structure is an impactful compromise between not defining any constraints and the materialization of constraints. The effort to create a PatchIndex is in the order of materialized views, while SortKeys and JoinIndexes require significantly more time to create. As PatchIndexes only add a single bit per tuple, they show a moderate memory overhead, which is only surpassed by the SortKey, which reorders data and avoids storing additional metadata. For distinct, and sort queries, the PatchIndex reaches a performance impact comparable to the specialized materialization approaches, which even holds for quite high exception rates. While update support is a drawback of materialized views and SortKeys, the PatchIndex approach offers lightweight support for update operations.

\section{Conclusion}\label{conclusion}
In this paper, we designed the sharded bitmap structure, which relies on virtually splitting an ordinary bitmap to keep update operations as local as possible to the affected memory regions. The data structure provides efficient support for insert, modify and delete operations while being scalable in size and adding only a small memory overhead of 0.39\% to the memory consumption of an ordinary bitmap. With the sharded bitmap as the underlying data structure, we designed the PatchIndex structure to maintain exceptions to arbitrary constraints, which allows the definition of approximate constraints in database systems. We described the integration of PatchIndexes for ``nearly unique columns'' (NUC) and ``nearly sorted columns'' (NSC) into the optimization of distinct, sort and join queries and provided mechanisms to efficiently maintain these constraints in the case of table updates, which avoid expensive index recreation and full table scans. In our evaluation, we compared the generic PatchIndex approach against materialized views, SortKeys and JoinIndexes as specialized materialization approaches in microbenchmarks as well as in the well-known TPC-H benchmark, proving that our approach is impactful for general purposes and even high exception rates.

Future work includes the discussion of additional constraints like approximate constancy of column values or arbitrary user-defined constraints that rely on the data semantics and their impact on query execution. Additionally, the concept of PatchIndexes offers possibilities for the field of approximate query processing \cite{li_approximate_2018}, as the PatchIndex contains information that hold for the major part of the data and therefore allows to generate approximate results on the whole dataset.
As the sharded bitmap allows to keep operations local to single shards, it offers the possibility for lightweight compression. Typically, bitmaps are compressed using run-length encoding, which could reduce the PatchIndex memory consumption especially for low exception rates. Although we presented an efficient mechanism to discover PatchIndex deltas in the case of updates by avoiding full table scans, further data structures like bloom filters or index structures could, if present in the database system, enhance the discovery of exceptions to approximate constraints caused by update operations.

\begin{acks}
This work was funded by the Actian Corp.
\end{acks}

\bibliographystyle{ACM-Reference-Format}
\bibliography{literature}

\end{document}